# Thermal shift of the resonance between an electron gas and quantum dots: What is the origin?

**Fabian Brinks, Andreas D. Wieck, and Arne Ludwig**[*]

*Lehrstuhl für Angewandte Festkörperphysik, Ruhr-Universität Bochum, D-44801, Germany*

[*]*corresponding authors e-mail address: arne.ludwig@rub.de*

The operation of quantum dots at highest possible temperatures is desirable for many applications. Capacitance-voltage spectroscopy (C(V)-spectroscopy) measurements are an established instrument to analyze the electronic structure and energy levels of self-assembled quantum dots (QDs). We perform C(V) in the dark and C(V) under the influence of non-resonant illumination, probing exciton states up to $X^{4+}$ on InAs QDs embedded in a GaAs matrix for temperatures ranging from 2.5 K to 120 K.

While a small shift in the charging spectra resonance is observed for the two pure spin degenerate electron s-state charging voltages with increasing temperature, a huge shift is visible for the electron-hole excitonic states resonance voltages. The $s_2$-peak moves to slightly higher, the $s_1$-peak to slightly lower charging voltages. In contrast, the excitonic states are surprisingly charged at much lower voltages upon increasing temperature. We derive a rate-model allowing to attribute and value different contributions to these shifts. Resonant tunnelling, state degeneracy and hole generation rate in combination with the Fermi distribution function turn out to be of great importance for the observed effects. The differences in the shifting behavior is connected to different equilibria schemes for the peaks – s-peaks arise when tunneling-in- and out-rates become equal, while excitonic peaks occur, when electron tunneling-in- and hole-generation rates are balanced.

## 1. Introduction

Self-assembled quantum dots (QDs) are promising candidates for future quantum technology applications as well as an interesting model system for studying fundamental quantum mechanics in a defined solid state environment. QDs can be used as suitable sources for indistinguishable single photons [1] for quantum communication [1,2,3], as qubits for quantum computing applications [4] or as repositories for electrons and holes, which is useful for creating new types of memory devices [5]. Also devices for energy conversion as thermoelectric energy harvesters [6] or QD solar cells [7] are envisioned. The investigation of the system's behaviour at increasing temperature is desirable for all those applications. That requires a deeper understanding of the charge carrier transfer processes, i.e. tunnelling. Thereby, new devices operating at higher temperatures than liquid helium are in reach.

The energy level structure of QDs can be revealed by C(V)-spectroscopy [8]. Warburton et al. modelled the different types of Coulomb-interaction in such systems and derived the corresponding energies theoretically [9]. Recently, it has been shown, that by creating holes in self-assembled QDs in a Schottky-diode under illumination, also excitons of different charge-state can be detected electrically [10,11]. Bayer and Forchel [12] performed temperature dependent investigations and determined the QD's homogenous linewidth increasing only by a tiny amount compared to $k_BT$, making QDs promising candidates for higher temperature applications. A study of the tunnelling dynamics is performed by Luyken et al. [13] and a description of the differences for in- and out-tunnelling rates due to degeneracy of the various electronic states were given by Beckel et al. [14].

However, a complete report of the tunnelling dynamics of self-assembled QDs coupled to an electron reservoir and the change with increasing temperature has not been given yet. We observe and provide an explanation for a huge thermal shift of excitonic states and a smaller shift for s-peaks in the gate voltage resonance for the C(V) of our QDs. We explain that shift of the peaks with electrons of energies disparate to the electrochemical potential $\mu_F$. The difference in the behavior of the s-peaks and the excitonic peaks is interpreted as a consequence of the different resonant tunnelling conditions: A resonant tunnelling in and out equilibrium for the s-peaks and a hole generation, tunnelling-in and recombination equilibrium for the excitonic peaks.

The manuscript is organized as follows. In section 2 the sample structure and the experimental details are described. Section 3.1 deals with the non-illuminated and section 3.2 with the illuminated measurements. Section 4 summarizes and concludes our findings. While the physics for understanding the experiments are motivated in the paper, in the supplementary material excessive additional information on the sample structure, experimental details, the theoretical model and more experimental data can be found.

**2. Sample structure and experimental details**

The experiments are performed on a layer of self-assembled InAs QDs grown in the Stranski-Krastanow growth mode [15] in a molecular beam epitaxy system. The layer is embedded in a MIS-structure consisting of a 300 nm thick highly n-doped ($n = 1.8 \times 10^{18}$ cm$^{-3}$) back contact, an undoped tunnelling barrier of $d_0 = 25$ nm thickness followed by the QD-layer. A 185 nm GaAs/AlAs superlattice prevents leakage currents and increases the efficiency of hole trapping. On top of that, a 20 nm semi-transparent gold gate allows the application of a voltage as well as the illumination of the structure with an LED driven by a current $I$. The complete structure of the sample is sketched in Figure 1 together with the corresponding band structure.

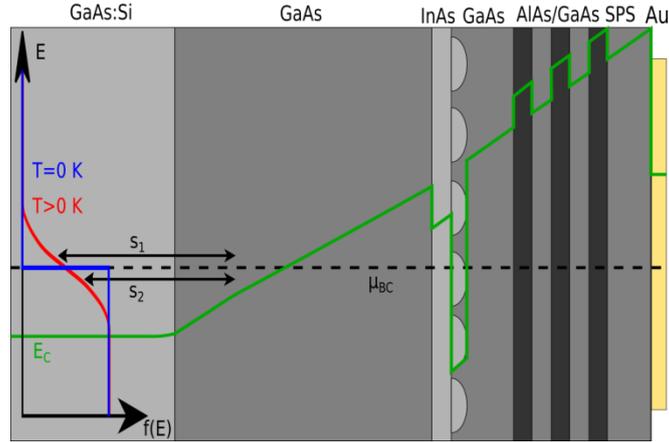

Figure 1: Sample structure together with the corresponding band structure (green) (not to scale). On the left hand side two different Fermi distribution functions are shown for T = 0 K (blue) and T > 0 K (red). The dashed line shows the position of the electrochemical potential in the back contact. The values 1/3 and 2/3 of the Fermi function are marked and indicate the resonance points for the $s_1$- and $s_2$-peaks.

The measurements are performed in a He-closed-cycle-cryostat with a base temperature of 2.5 K. The temperature of the sample can be controlled in the range of 2.5 K to 300 K via heating. For performing C(V)-spectroscopy, an AC-voltage $U_{AC} = 20$ mV$_{eff}$ is added to a DC-gate-voltage, which results in the charging and discharging of the capacitor consisting of the back contact and the gate. The 90° phase shifted charge current is measured using a lock-in-amplifier. When electrons in the back contact come into resonance with the energy levels of the QD, they are able to tunnel into the dot, which leads to an increased charging current and therefore a local maximum in the C(V)-spectrum. The applied gate voltage can be converted into an energy using the geometrical lever arm approach [16] relating the total length between back contact and gate $d_{tot}$ to the nominal tunnel length $d_0$. In our sample this quantity is $d_{tot}/d_0 = 8.4$.

### 3. Results and interpretation

**3.1 s-peaks:**

In Figure 2 (a) one can see the spectra of measurements at non-illuminated (dark) conditions around the position of the s-peaks. Those two charging peaks originate from electron charging of the QD ensemble with the first and second electron on an s-shell. The peaks are separated by the Coulomb repulsion, lifting the spin degeneracy and broadened due to the ensemble inhomogeneity, i.e. the QDs are not identical and get charged at slightly different energy values. The peaks broaden even more with increasing temperature. A fluctuation of the QDs Eigen energies due to interaction with the phonon bath leads to a temperature dependent broadening of up to 0.1 meV at 100 K [12], certainly not enough to explain the observed effect. We attribute the main contribution to the smeared Fermi distribution in the back contact. The charging state of a QD thus underlies a certain statistical probability with an effective width of $k_BT$.

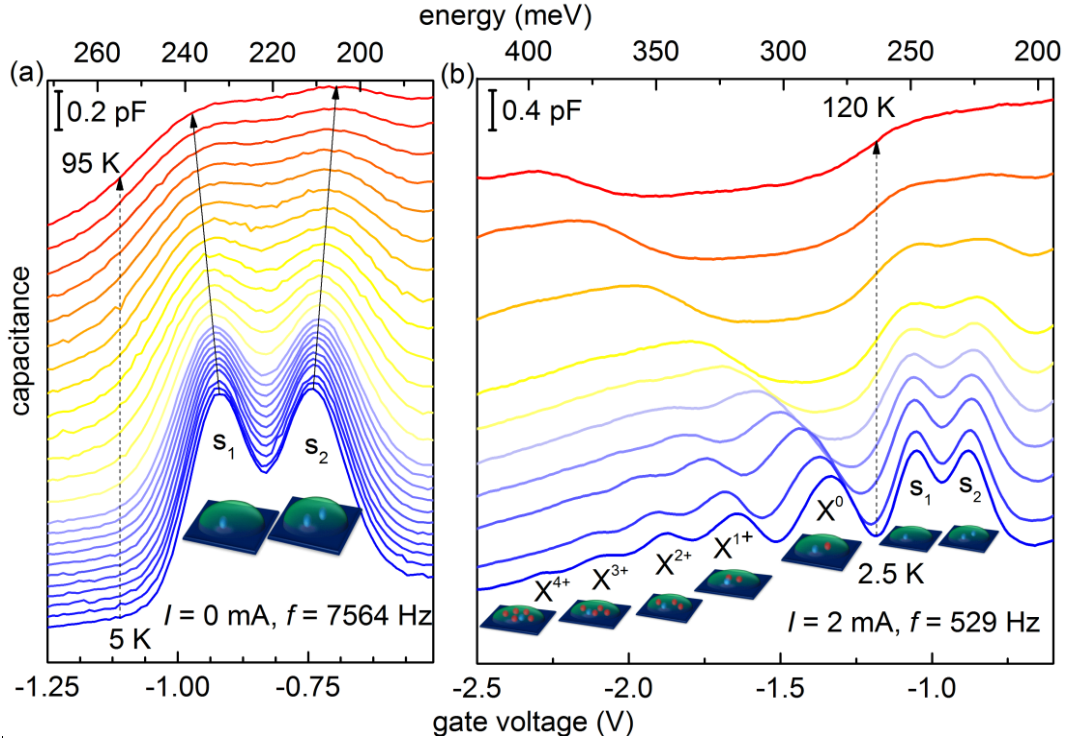

Figure 2: C(V)-spectra (a) around the position of the s-peaks for a non-illuminated sample at $f = 7564$ Hz, where the solid arrows indicate the positions of the s-peaks, and (b) around the position of the excitonic peaks for an illuminated sample with $I = 2$ mA and $f = 529$ Hz. The spectra are offset vertically by 0.02 pF/K (a) and 0.03 pF/K (b), with respect to the C(V) spectra at lowest temperature for clarity. The energy axis is calculated by the simple lever approach, the blue and red spheres in the schematic quantum dots represent electrons and holes.

Looking more carefully to the spectra, it appears that upon temperature rise the $s_1$-peak shifts to lower voltages, while the $s_2$-peak does the opposite (solid arrows in Figure 2 (a)). To explain this, we will first exclude any significant peak-shift due to Coulomb repulsion between the electrons from electrostatic origins. Then, we apply our model.

An outward peak shift could be explained by an unpredicted increased Coulomb repulsion. Parameters that change with temperature are the dielectric constant (increase [17]), the lattice parameter (increase [18]), the effective mass (decrease [19]), the confinement energy (barely any change) and the bandgap energy (decrease [20]). In the Coulomb integral [9], the dielectric constant is in the denominator and would thus lead to a reduced Coulomb repulsion. A larger lattice parameter or/and a lower effective mass would extend the electron wave functions and the outcome is the same. The peaks would shift inwards. A change in the carrier binding to first approximation would move both peaks parallel, which is not observed. As we restrict our observation to the conduction band, a change in the bandgap would lead to a parallel shift, if the Schottky barrier height is altered. As the peaks move outwards and do not shift parallel or move inwards, we can disclaim all these hypotheses. Our explanation of the observed effect includes a consideration of the tunnelling rates. The resonance condition, *i.e.* the peak in the

C(V)-spectrum occurs, if there is an equilibrium between the respective charge states $m$ in the QDs. A simple argument is the following: The transition and thus peak called $s_1$ occurs when $m$ changes from 0 to -1. Thus the probability to find both charge states is identical and equal to 1/2. To fulfil this, the tunnelling rates into the QD and tunnelling rates back to the reservoir have to be equal.

$$\Gamma_{in} = \Gamma_{out} \tag{1}$$

The rates for the tunnelling processes from a charge state $\widetilde{m}$ into a charge state $m$ at a certain energy are given by:

$$\Gamma^{e^-}_{\widetilde{m} \to m} \propto g_{\widetilde{m} \to m} T_{\widetilde{m} \to m}(E) f(E,T) D(E), \quad \widetilde{m} > m \tag{2}$$

for electron tunnelling-in, and

$$\Gamma^{e^-}_{\widetilde{m} \to m} \propto g_{\widetilde{m} \to m} T_{\widetilde{m} \to m}(E) [1 - f(E,T)] D(E), \quad \widetilde{m} < m \tag{3}$$

for electron tunnelling-out. $g$ is the degeneracy of the final state in the QD, $T(E)$ is the tunnelling probability, $f(E,T)$ is Fermi's distribution function and $D(E)$ is the density of states in the back contact. The first two terms on the right hand side describe the tunnelling probability for one electron. The latter two factors in the product give the density of electrons in the back contact at a certain energy $E$ for $\Gamma_{in}$, or the density of free states in the back contact for $\Gamma_{out}$, respectively.

For the $s_1$-state, an electron with arbitrary spin orientation can tunnel into the dot, corresponding to a twofold degeneracy ($g^S_{0 \to -1} = 2$), while if the dot is occupied, there is either a spin-up or a spin down electron in the QD, which means no degeneracy, i.e. $g^S_{-1 \to 0} = 1$. The peak condition (1) at a resonance energy $E^{s_1}_{res}$ together with the in- and out-tunnelling rates (2) and (3) is solved by:

$$f(E^{s_1}_{res}, T) = 1/3 \tag{4}$$

Calculating the resonance energy peak shift for this value of the Fermi distribution yields:

$$E^{s_1}_{res} - \mu_F = \ln(2) \, k_B T := m_1 T \tag{5}$$

For the $s_2$-state the QD is already filled with an electron of a certain spin orientation. According to that, only an electron with opposite spin may enter the dot and tunnelling-in is not twofold spin degenerated ($g_{-1 \to -2} = 1$). The two electrons in the dot are now energetically degenerated and thus there are two possibilities of spin directions leaving the dot. Therefore, tunnelling-out is twofold spin degenerated ($g_{-2 \to -1} = 2$). This leads to:

$$f(E^{s_2}_{res}, T) = 2/3 \tag{6}$$

$$E^{s_2}_{res} - \mu_F = -\ln(2) \, k_B T =: m_2 T \tag{7}$$

This is an important finding, as for both states $s_1$ and $s_2$ the resonance condition is not fulfilled for the position of the Fermi energy $\mu_F$ at $f(E) = 1/2$, as one could expect. Instead, the resonant points shift to higher energies for $s_1$ and lower energies for $s_2$ due to the difference in their tunnel-in and -out degeneracy. This corresponds to lower respectively higher gate voltages. It is worth to mention that the resonance peak shift is independent of material parameters and purely a consequence of the Fermi distribution and the level of degeneracy in the QDs. It resembles the entropy of a two level system.

The observed and calculated shifts are shown in Figure 3 (a). The slopes calculated according to the simple lever arm approach [16] are $m_1 = 0.998 \times \ln(2) k_B$ for $s_1$ and $m_2 = -0.919 \times \ln(2) k_B$ for $s_2$, which is in good agreement with our model. The tendency to smaller slopes for the $s_2$-peak, especially at higher temperatures is qualitatively explained by the changes of the electrostatic environment changing the Coulomb repulsion (see disclaimed hypotheses above).

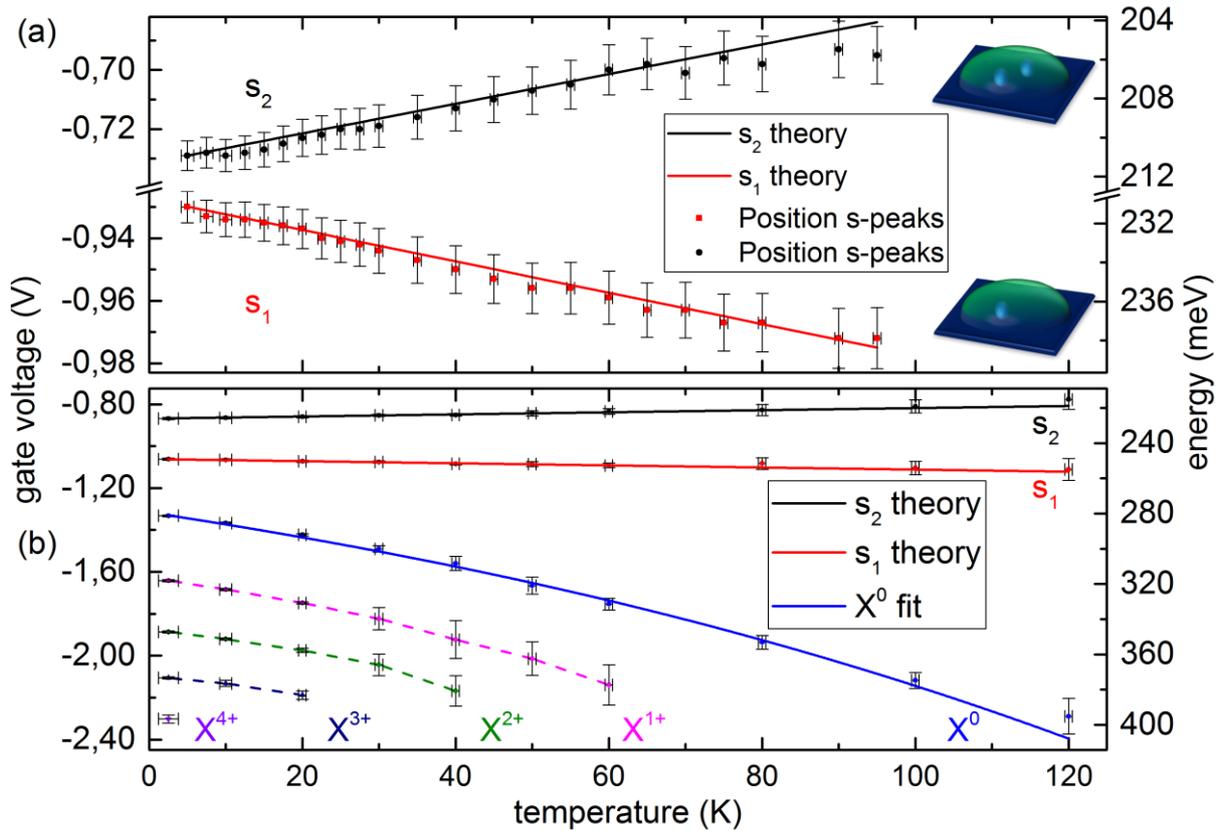

Figure 3: Peak positions as a function of temperature. The energy axes are calculated by the simple lever approach. (a) s-peaks in the non-illuminated case for $f = 7564$ Hz. The solid lines correspond to the model according to equations (5) and (7). (b) Excitonic peaks and s-peaks in the spectra of an illuminated sample at $I = 2$ mA and $f = 529$ Hz. The dashed lines are guides to the eye, the solid lines correspond to theoretical shifts according to equations (5) and (7) for the s-peaks and a fit to the resonance condition equation (8) for the $X^0$ peak with a hole generation rate of 400 Hz.

### 3.2 Excitonic-peaks:

The spectra of excitonic charging peaks for different temperatures are shown in Figure 2 (b). The sample has been illuminated with an LED current of 2 mA and the spectra have been measured with an AC-frequency of 529 Hz. The positions of the peaks are plotted in Figure 3 (b).

Such an excitonic peak appears at low temperatures, when the Fermi energy is aligned to the electron Eigen energy in the QD charged by a number of holes via illumination [11]. These Eigen energies originate from attractive Coulomb interaction of the holes and the electron and thus appear at lower gate voltages than the s-charging peaks.

With increasing temperature, all excitonic charging peaks are shifted to lower voltages. The shift is huge compared to that of the s-peaks. For $X_0$ the variation is nearly 1 V of gate voltage in the range from 2.5 to 120 K, whereas both s-peaks shift for ~40 mV in this temperature range – a difference of more than one order of magnitude.

We attribute the main reason to a different peak creation mechanism. For the excitonic peaks, only in-tunnelling of electrons is relevant. The tunnelling-out processes, which counteract the peak shifting in case of $s_1$, or even invert it for $s_2$, do usually not appear, as the recombination time of an electron hole pair is on the order of one nanosecond [3], orders of magnitudes faster than the AC period of ~2 ms in our experiment.

In contrast to the tunnelling-out process of the s-peaks, an excitonic charging peak can only occur, if a hole is generated in the QD. This generation is the equivalent of an electron out-tunnelling event. The resonance condition is fulfilled, when the electron tunnelling-in rate $\Gamma^{e^-}_{1 \to 0}$ for one electron tunnelling into a single positively charged QD and the hole generation rate $\Gamma^{h,\,gen}_{0 \to 1}$ where one hole is generated from an uncharged QD, are equal:

$$\Gamma^{e^-}_{1 \to 0} = \Gamma^{h,\,gen}_{0 \to 1} \qquad (8)$$

The hole generation is a complicated process. Electron hole pairs can be created in different parts of the sample: The tunnelling barrier, the wetting layer or directly in the QDs. A portion of the generated electron hole pairs will certainly directly recombine. However a significant amount of electrons drift, diffuse or tunnel into the back contact, while the holes relax into the QDs. A stronger illumination creates more electron hole pairs, while a more negative gate voltage makes it easier for electrons to leave the QDs. A precise expression for the rate is unknown. As we expect no strong change with applied gate bias, for our approximation we restrict our analysis to a constant hole generation rate.

The tunnelling in rate of electrons for the excitonic recombination $\Gamma^{e^-}_{1 \to 0}$ might originate from the following three contributions:

1. Electrons entering the QDs above the edge of the tunnelling barrier conduction band
2. Spatially indirect recombination of an electron in the back contact with a hole in the QD [21]
3. Resonant tunnelling in QD states

Mechanism 1 should be negligible, because the energy to overcome is typically 250 meV and thus the number of electrons beyond the conduction band edge is low at the investigated temperatures.

The second mechanism to be considered is the spatially indirect recombination. Electrons in the back contact and holes in the QDs are quantum mechanical particles represented by wave functions. If there is an overlap in the wave functions, recombination can take place according to Fermi's golden rule. That means electrons in the back contact can recombine with holes without having a resonant state in the QD. This recombination is usually much slower than spatially direct recombination [21]. In our case, it might play a crucial role as this time might become comparable to the inverse tunnel rate. We develop a rate equation (see supplementary) to attribute this annihilation process.

$$\Gamma_{1\to 0}^{e^-,\text{indirect}} \propto f(E,T)|\langle \Psi_{BC}^e|e_{12}|\Psi_{QD}^h\rangle|^2 \qquad (9)$$

where the terms of the dipole matrix element are the dipole strength $e_{12}$, the electron wavefunction in the back contact $\Psi_{BC}^e$ and the hole wavefunction in the QD $\Psi_{QD}^h$. An analysis of the gate voltage dependency yields a smooth bias dependency of the rate (see Figure 3, supplemental material). Labud et al. [11] show, that the excitonic peak-position is barely hole generation rate dependent by varying the illumination intensity nearly two orders of magnitude, while the smooth bias depence would imply such a dependency. Thus this spatial indirect recombination has to be less important and another effect with a steep increase in rate must be responsible for the observed excitonic peak and its temperature shift.

Effect 3, the resonant tunnelling, only happens, when an electron is resonant to a QDs energy level. The electron tunnels resonantly into the dot, relaxes to the ground states energy level within a few picoseconds [22] (if not already tunnelled into the ground state in the first place) and recombines with the hole. The recombination takes place in a time of a few nanoseconds [3], much faster than the tunnelling time of the electron in the range of a few microseconds [13]. Therefore, resonant tunnelling is the limiting process and recombination happens efficiently and fast afterwards.

The tunnel in rate for the resonant tunnelling-in hole occupied states is given by an expression formed by contributions of resonant tunnelling into s-, p-, d-, f-,…states and the corresponding degeneracies, also see supplementary for a derivation and plot.

$$\Gamma_{\widetilde{m}\to m}^{e^-} \propto f(E,T)D(E)\sum_{j=s,p,d,\ldots} g_{\widetilde{m}\to m}^j T_{\widetilde{m}\to m}^j(E), \qquad \widetilde{m} > m \qquad (10)$$

A steep increase of several orders of magnitude is found at low temperatures whenever a level comes in resonance with the electrochemical potential of the back contact. This steep increase certainly gives rise to a sharp peak in the C(V)-spectra at practically constant bias condition even for a large variation in hole generation rate. In other words: The equation $\Gamma_{1\to 0}^{e^-} = \Gamma_{0\to 1}^{h,\text{gen}}$ and thus the peak condition in the C(V)-spectra is fulfilled for a wide range of hole generation rates at nearly the same bias condition. With increasing temperature, the steep slope smears out due to contributions of electrons tunnelling into the QDs at higher energies and with a much shorter tunnel barrier, explaining the shift in the peaks (Figure 3). The discrepancy of our simple model to our experimental findings at temperatures above 80 K might be a consequence of a larger hole generation rate due to a higher electron out-tunnelling rate during the generation process. Neglected before, electron tunnelling-out directly after tunnelling-in without electron-hole annihilation might happen as well, if the electron tunnelling-out rate at high reverse biases becomes the same order of magnitude as the recombination rate in the QD.

Overall, our model reproduces well the large $X^0$ peak shift of exciton annihilation observed in the temperature dependent C(V)-spectroscopy experiment.

This finding is applicable to other fields where a charge reservoir is coupled to a QD and elevated temperatures are desired, like thermoelectric energy converters, QD solar cells or electrically driven or electrically controlled deterministic solid state single photon sources. The specifications for energy barriers needed to operate the latter at elevated temperatures will be challenging. In our experiment we used a triangular shape barrier, promoting the observed severe excitonic energy shifts, as the thermally excited carriers involved in the resonant tunnelling process are much closer to the QDs (see Figure 1). For rectangular barriers, the length of the tunnel barrier would not change with applied bias and the effect is thus anticipated to be much less pronounced. Other options come along with energy filters in the barrier by a Type-III broken gap band line up, quantized states and k-space overlap engineering [23].

As the observed s-state shifts relies on fundamental constants only, a metrology method for determining $k_B$ or use as a sensitive thermometer is possible.

## 4. Summary and conclusions

Self-assembled QDs are investigated by C(V)-spectroscopy under illumination and various temperatures. We observe thermal shifts for s-states as well as for excitonic states. The shifts are a consequence of the change in Fermi's distribution function and resemble the entropy of a two-level system. For the s-states, we found the differences in the degeneracies of tunnelling-in and -out of the QDs to be a crucial factor. In contrast to intuition, the resonant tunnelling condition does not occur at a value of the Fermi distribution function of $f(E,T) = 1/2$, but at $f(E_{\text{res}}^{s_1}, T) = 1/3$ (and $f(E_{\text{res}}^{s_2}, T) = 2/3$) for the $s_1$ (and $s_2$) state, in good agreement with our measured values. Especially, the $s_2$-peak shifting unexpectedly opposite to the $s_1$ peaks could be understood. A significantly stronger shift for excitonic peaks was explained by a different mechanism. After electron tunnelling-in, tunnelling-out as a counteracting factor cannot take place due to a fast recombination process. In contrast, the hole generation rate is the counteracting process here with a completely different energy dependency from the tunnelling processes, yielding the observed strong shift. The rate equations for the peak shifts could be deduced from an overlying master equation, which gives a complete picture.

The data shows that the occupancy of the eigenstates in the QDs is also understood at temperatures exceeding liquid nitrogen, extending their application potential in various fields as energy harvesting, metrology and quantum information technologies. Also limitations and the great challenge towards electrically driven single photon sources operated at elevated temperatures are quantifiable by our findings.

**Acknowledgement**

We would like to thank Nadine Viteritti for expert sample preparation and Patrick A. Labud for fruitful discussions in the beginning of the project. We gratefully acknowledge support from BMBF Q.Com-H 16KIS0109 and DFG-TRR160.

**Author contributions**

F.B. and A.L. performed the experiment, interpreted the data and wrote the manuscript. A.L. grew the heterostructure, developed the model and led the project. A.D.W. supervised the research. All Authors contributed to the manuscript and discussed it to its final form.

# Supplementary information to:
# Thermal Shift of the Resonance between an Electron Gas and Quantum Dots: What is the Origin?

Fabian Brinks, Andreas D. Wieck, and Arne Ludwig

*Lehrstuhl für Angewandte Festkörperphysik, Ruhr-Universität Bochum, Bochum, 44780, Germany*

## Section I: Additional experimental details

The central layer sequence is already given in the main manuscript. The entire layer sequence is presented in Table 1.

Table 1: Heterostructure layer sequence starting with the substrate and ending with the surface (10 nm GaAs oxidation protection cap). Pyrometrically measured surface temperatures are mentioned in the text below.

| Layer | Thickness (nm) | |
|---|---|---|
| GaAs | 50 | |
| AlAs | 30 x | 2 |
| GaAs | | 2 |
| GaAs | 50 | |
| GaAs:Si | 300 | |
| GaAs | 25 | |
| InAs | 0.45 | |
| GaAs | 8 | |
| GaAs | 3 | |
| AlAs | 41 x | 3 |
| GaAs | | 1 |
| GaAs | 10 | |

The semiconductor growth was performed on an Epineat molecular beam epitaxy (MBE) machine on an epiready vertical gradient freeze (VGF) GaAs (001) $\pm$ 0.1° semi-insulating wafer.

The base pressure is in the range of $10^{-10}$ mbar and maximum achieved mobilities within the growth campaign were $3.5 \times 10^6$ cm$^2$/Vs for a single heterojunction modulation doped two dimensional electron gas at $T = 4.2$ K.

The arsenic beam equivalent pressure was $9.6 \times 10^{-6}$ torr and $6.8 \times 10^{-6}$ torr measured with a flux tube placed into the molecular beam path in front of the substrate.

The growth rates were adjusted before the sample growth by reflection high-energy electron diffraction (RHEED) oscillation measurements to be $r_{\text{GaAs}} = 0.1996$ nm/s, $r_{\text{AlAs}} = 0.1034$ nm/s and $r_{\text{InAs}} \approx 0.075$ nm/s.

The wafer temperature was measured with a pyrometer. The heterostructure was grown at $600\,°\text{C}$ with exception for the QDs, $1.7$ Monolayers InAs, grown at $525\,°\text{C}$, a QD capping layer of $8$ nm GaAs grown at $500\,°\text{C}$ and additional $3$ nm GaAs grown with increasing temperature. After another break of $80$ s to increase the temperature back to $600\,°\text{C}$, the short period superlattice growth was started.

The bandstructure of the processed Schottky diode is shown in Figure 1 along with a sketch of the heterostructure layer sequence.

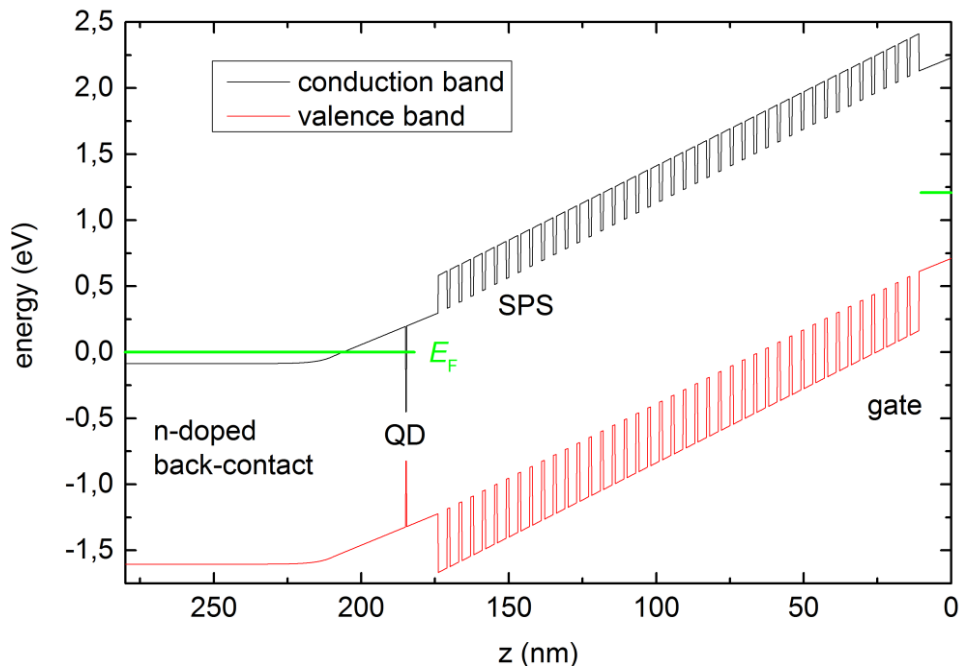

Figure 1: Band structure of the uncharged structure at a gate bias of -1.2 V. The QDs are represented by a small quantum well in this one-dimensional band structure simulation. The SPS is a short period superlattice, blocking holes from leaving the QDs.

Ohmic contacts are processed to a $3.5 \times 5$ mm$^2$ sample by indium solder to the corners. This is sufficient to create an ohmic contact to the buried n-doped back-contact. Semitransparent $20$ nm gold gates with lateral dimensions of $300 \times 300$ μ$m^2$ are processed by standard photolithography (Karl Suss mask aligner and Shipley Microposit 2000/ SP2510 photoresist) and thermal evaporation from a tungsten crucible under high vacuum conditions.

After lift-off, removing the photoresist and gold on the latter, the sample is glued to a 16 pin dual inline chip carrier and gates and ohmic contacts are ultrasonically wedge-wire bonded with $25\,\mu\mathrm{m}$ diameter Al-wires.

The sample was cooled in a pulse tube or also called closed cycle cryostat (CCC) with short circuit between the ohmic contact and the gate.

The sample leads are made from stainless steel to have low temperature impact in the cold stage of the cryostat. Cryogenic temperature measurements are done with a temperature sensor, being in a temperature regulation loop with a Lake Shore 332 temperature controller and electric heater at the copper cold finger of the cryostat.

Measurements are done with a Zurich instruments MFLI Lock-In amplifier using the internal analog circuit to add a sine voltage with $U_{\mathrm{AC}} = 20\,\mathrm{mV}$ to the Aux-DC-voltage ($U_{\mathrm{Gate}}$) scanned to perform the spectroscopy.

The capacitive $90\,°$ phase shift current signal from the ohmic contact is recorded by the demodulator of the Lock-In amplifier.

Non-illuminated capacitance-voltage (C(V)) spectra are recorded at an AC-frequency of $7564\,\mathrm{Hz}$. At this AC-frequency, no suppression of the capacitance peaks due to incomplete tunnelling processes could be observed (see figure 2 (a) in the main text). An integration time of $T_{\mathrm{c}} = 10\,\mathrm{ms}$, a filter of $48\,\mathrm{dB}$ and a wait time of $21 \times T_{\mathrm{c}}$ was chosen to obtain sufficiently smooth data.

To reduce a suppression of the tunnelling process from too high frequencies, for the illuminated measurements a frequency of $529\,\mathrm{Hz}$ was chosen [1]. An integration time of $T_{\mathrm{c}} = 80\,\mathrm{ms}$, a filter of $6\,\mathrm{dB/oct}$ and a wait time of $2 \times T_{\mathrm{c}}$ was set to obtain sufficiently smooth data, an averaging of 16 measurements per gate voltage has been performed with the Lock-In amplifier. For temperatures T = 2.5 .. 60 K, each measurement has been done twice and averaged afterwards for a further reduction of noise influence. For T = 80 K, there have been 3 repetitions and 4 for T = 100 K and 120 K because of increasing noise. A Savitzky-Golay smoothing filter (10 points) was applied to the data in Figure 2 (b) of the main text. Illumination of the sample is performed by an LED emitting at a centre wavelength of $\lambda \approx 919\,\mathrm{nm}$ at $T = 4.2\,\mathrm{K}$ driven by a constant current source-measure unit (Keithley 236).

## Section II: Modelling the C(V) spectra
## Section II a): The master equation

To model the observed QD-levels, a harmonic oscillator model [2], Coulomb interactions [3] and atomistic pseudopotentials [1] are successfully applied. To explain the C(V)-spectra at low temperatures, the following simple argument is used: When electrons in the back contact come into resonance with the energy levels of the QD, they are able to tunnel into the dot, which leads to a decreased width of the dielectric, thus an increase of the capacitance, thus an increased charging current and therefore a local maximum in the C(V)-spectrum [4]. Taking effects of level degeneracy in the QDs [5] and elevated temperatures into account, a master equation [6] describing the charge dynamics in QDs is used.

$$\dot{p}_m = \sum_{\widetilde{m}\neq m} \Gamma_{\widetilde{m}\to m} p_{\widetilde{m}} - \sum_{\widetilde{m}\neq m} \Gamma_{m\to\widetilde{m}} p_m \tag{1}$$

where $m$ is the charge state of the QD (e.g. $m = -1$ for one electron, $m = 1$ for one hole) and its probability $p_m$. $\Gamma_{\widetilde{m}\to m}$ is the rate of carrier transfer between back contact and QD from the QD's charge state $\widetilde{m}$ to $m$, typically a tunnel rate. In the quasi-equilibrium of C(V)-measurements, the temporal variance of the state probability vanishes, i.e.

$$\dot{p}_m = 0. \tag{2}$$

Further,

$$\sum_m p_m = 1. \tag{3}$$

In our model we will show, that the exciton annihilation rate is basically an equilibrium between the hole generation rate $\Gamma_{m\to\widetilde{m}}^{\text{h, gen}}$ on the one hand and electron tunnel in rate $\Gamma_{\widetilde{m}\to m}^{e^-}$. We consider tunnelling in various shells, level relaxation/thermalization $\Gamma_m^{e^-,\,\text{relax}}$ (ignored for tunneling into an s-state) and electron-hole recombination $\Gamma_m^{X^m,\,\text{rec}}$. The latter three (electron tunnelling in, thermalization and recombination) can be seen as a series process. Level relaxation/thermalization $\Gamma_m^{e^-,\,\text{relax}}$ and electron-hole recombination $\Gamma_m^{X^m,\,\text{rec}}$ happens on timescales of picoseconds and nanoseconds respectively. As electron tunnelling $\Gamma_{\widetilde{m}\to m}^{e^-}$ happens on a microsecond timescale, we can approximate this series of processes as

$$\left(\frac{1}{\Gamma_{\widetilde{m}\to m}^{e^-}} + \frac{1}{\Gamma_m^{e^-,\,\text{relax}}} + \frac{1}{\Gamma_m^{X^m,\,\text{rec}}}\right)^{-1} \approx \Gamma_{\widetilde{m}\to m}^{e^-}. \tag{4}$$

In parallel, also spatial indirect tunnel recombination $\Gamma_{\widetilde{m}\to m}^{e^-,\,\text{indirect}}$ is considered to make the picture whole.

The rates in this equation are motivated in the following sections.

## Section II b): Electron tunnel rate equation

The rate equations for electron tunnelling (equation (2) and (3) in the main text) in the context of the master equation are:

$$\Gamma^{e^-}_{\tilde{m}\to m} \propto f(E,T)\Theta(E) \sum_{j=s,p,d,\dots} g^j_{\tilde{m}\to m} T^j_{\tilde{m}\to m}(E), \qquad \tilde{m} > m \tag{5}$$

for electron tunnelling in the QD, where the charge state is more negative after the tunnelling process ($\tilde{m} > m$) and

$$\Gamma^{e^-}_{\tilde{m}\to m} \propto (1-f(E,T))\Theta(E) \sum_{j=s,p,d,\dots} g^j_{\tilde{m}\to m} T^j_{\tilde{m}\to m}(E), \qquad \tilde{m} < m \tag{6}$$

for electron tunnelling out.

The elements of these equations are the Fermi distribution $f(E,T)$ of the electron reservoir in the back contact, a theta distribution to account for the conduction band minimum in the back contact, the degeneracy of the $j$-th final state in the QD $g^j_{\tilde{m}\to m}$ and the tunnel probability into the $m$-th charge state in the $j$-th level $T^j_{\tilde{m}\to m}(E)$.

We will now derive the tunnel probability. Therefore we define the energy $E$ of the tunneling charges by setting $E = 0$ at the bottom of the conduction band in the doped region (see Figure 2). We can thus count the energy in the Fermi distribution starting from zero with an electrochemical potential of $\mu_F = 80$ meV for our $n = 1.8 \times 10^{18}$ cm$^{-3}$ doped back contact. We account for the conduction band energy of various QD-states from the conduction band edge using the harmonic oscillator potential approximation[2] and Coulomb interactions [1,3]. For the electron energy states we thus find energy levels $E^j_m$ with $m$-th charge state and $j$-th level. For one electron, $m = -1$ and $j$ are the s-, p-, d-,… single electron conduction band states. Coulomb interaction reduces (increases) these energy levels for additional holes (electrons). Please note that the $E^j_m$ are ionization energies from the QDs level to the GaAs-conduction band edge at the position of the QDs.

Using the WKB approximation, the expression for a triangular tunnel barrier is given by

$$T(E, d_{\text{tunnel}}) = exp\left(-\frac{2}{\hbar} \int_0^{d_{\text{tunnel}}} \sqrt{2m^*(V(z) - E)} dz\right) \tag{7}$$

The electrons (effective mass $m^*$) from the back contact have to tunnel into these energy levels through a barrier $V(z)$ that we approximate in the following to be triangular (We take a simple lever approach [7]). Thus the barrier effective height $(V(d_{\text{tunnel}}) - E)$ is given by these $E^j_m$ values for resonant tunnelling. The tunnel length $d_{\text{tunnel}}$ for each level is given by the horizontal distance between the QD and the conduction band edge and changes by simple geometrical relations with the applied gate voltage, which corresponds to an energy $E$ at the position of the quantum dot.

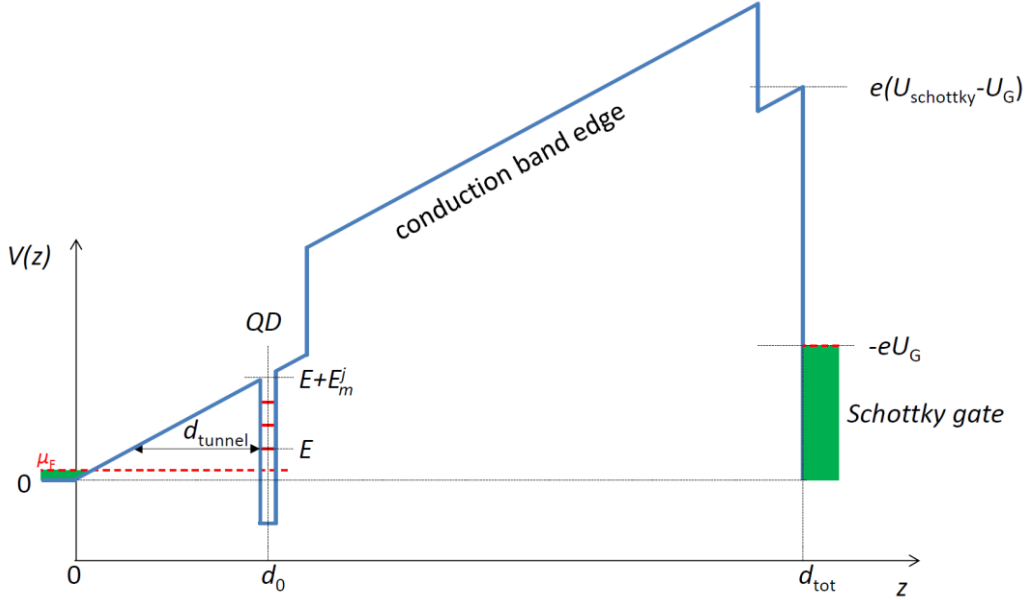

Figure 2: Simplified conduction band sketch. Electrons of the energy $E$ may tunnel through the barrier with a tunnel length $d_{\text{tunnel}}$ into the energy levels with ionization energy $E_m^j$. Band bending due to space charge regions is ignored in the simple lever approach.

$$d_{\text{tunnel}} = d_0 \left( \frac{1}{1 + \frac{E}{E_m^j}} \right) \tag{8}$$

Where $d_0$ is the nominal tunnel distance, 25 nm in our device.

The energy $E$ can be calculated using the following equation:

$$e(U_{\text{Schottky}} - U_G)\frac{d_0}{d} = E + E_m^j, \tag{9}$$

where $e$ is the elementary charge, $eU_{\text{Schottky}}$ is the Schottky barrier height (1.03 eV for Au on GaAs) and $U_G$ is the gate voltage. For gate voltages from $-2.8$ V to $-0.3$ V, equation (7) is solved for electron tunnelling in from charge state $\tilde{m} = 1$ (one hole) to $m = 0$ and plotted as a function of the gate voltage for different temperatures in Figure 3.

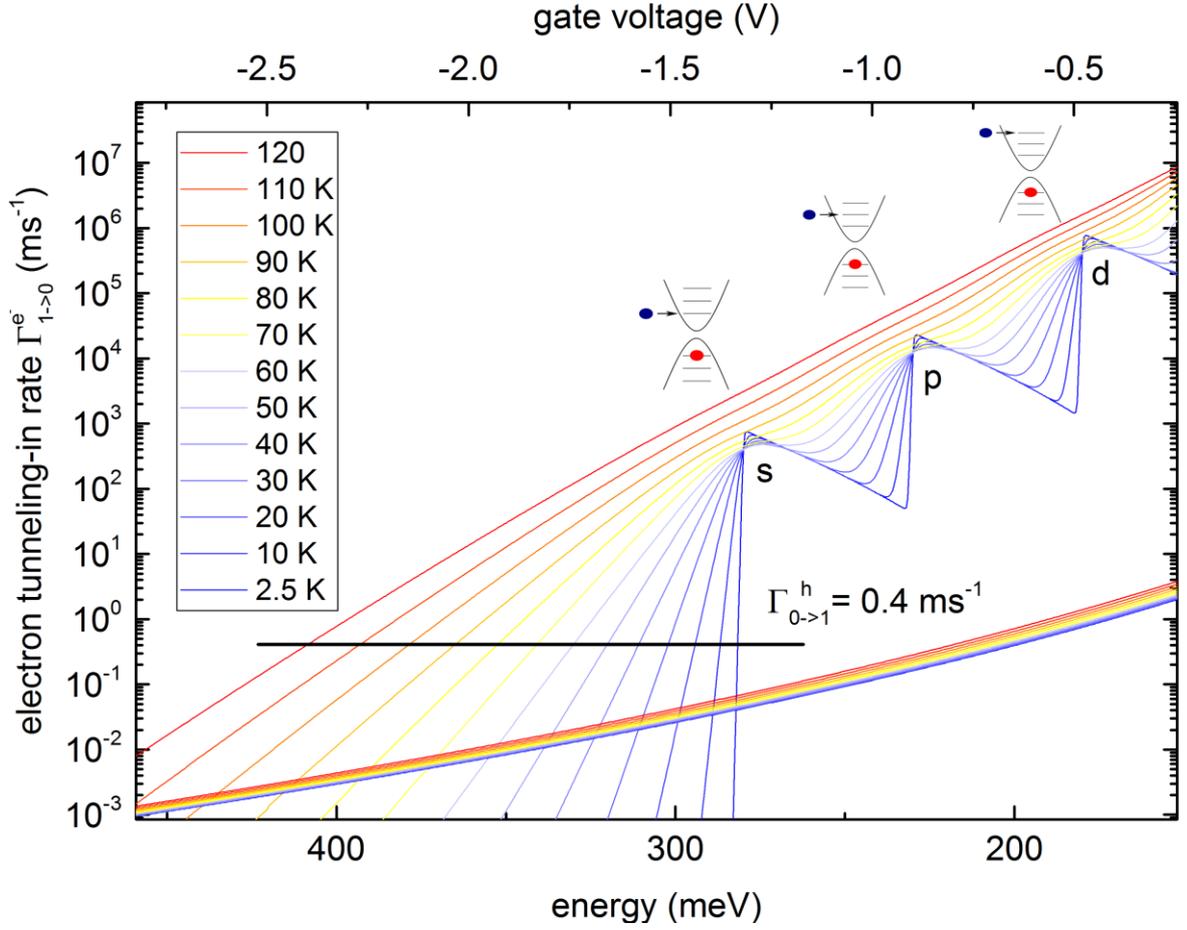

Figure 3: Calculated electron tunnelling-in rate for the $X^0$-state as a function of the gate voltage for different temperatures. The horizontal line marks the hole generation rate according to our data fit. The energy axis corresponds to $E + E_m^j$ (equation (9)). In contrast to resonant tunnelling (upper curves), a smooth increase is visible in the spatially indirect recombination rate (lower curves). The absolute rate values are estimated values adapted to our samples after Luyken et al. [4] and Rai et al. [8] (see section II e).

At low temperatures, steep increases are seen due to resonances of the QD levels with the electrochemical potential of the back contact. The continuous drop after each step originates from the fact that we take the modified tunnel barrier thickness at lowered band alignments into account. The higher step size for the higher levels arise partially from the larger degeneracy in the latter, making a larger number of states accessible for electron tunnelling in. Additionally, the tunnel coupling gets better due to the lower energy barrier and the shorter tunnel length. (Note: A step down of $\approx 1 \text{ ms}^{-1}$ is expected as the QD $E_0^s$-level falls below the conduction band edge at $U_G \approx -0.6$ V in the back contact, accounted by the theta-function in the tunnel probability equation effectively setting the electron density to zero at the lower conduction band edge. However, this step is not visible in the logarithmic representation of Figure 3). At higher temperatures, the steps transfer to a smoother increase due to the smeared out Fermi distribution function.

## Section II c): Indirect electron-hole recombination

We now want to derive an equation with which we can describe spatially indirect tunnel recombination [8], where an electron of the back contact at energies resonant or not resonant with the QDs energy level recombines spatially indirect with a hole in the QD.

Again we consider the potential barrier to be triangular and use the same arguments as for the resonant tunnelling process. The difference is that the sum of discrete resonance energies in equation (5) is replaced by an integral. The recombination rate, proportional to the dipole matrix element, is given by Fermi's golden rule where the electron-hole overlap and the dipole strength are considered.

$$|\langle \Psi_{\mathrm{BC}}^{\mathrm{e}} | e_{12} | \Psi_{\mathrm{QD}}^{\mathrm{h}} \rangle|^2 \tag{10}$$

As the dipole strength does not vary strongly, because the relative change in $d_{\mathrm{tunnel}}$ and the energy differences for the emitted photons are small, we take it as approximately constant. We consider the electron-hole overlap proportional to the Airy-function $Ai(E)$ of an electron at the locus of the QD, which is the solution of Schrödinger's equation for a triangular potential. As the WKB approximation used for the resonant tunnelling states yielded nearly identical results for this triangular potential wall, this is a good approximation and a great simplification. Please note that we should also consider the probability amplitude extend of the hole in the QD up to the electrons in the back contact. This is neglected however, as the high effective mass of the heavy hole state forces the hole-wave function quickly to zero. The spatially indirect tunnel recombination thus reads

$$\Gamma_{\widetilde{m}\to m}^{\mathrm{e}^-,\,\mathrm{indirect}} \propto \int_0^{E+E_m^S} f(E') Ai(E')^2 dE' \tag{11}$$

This function is plotted in Figure 3 lower curves, as a function of gate voltage and temperature with a guessed absolute value following the arguments below. The indirect recombination time is not known for our structures. However, varying the hole generation rate as done in section III would lead to a qualitative completely different behaviour as observed it in our experiment, if the rate would be significantly high. The observed exciton peak positions are barely hole generation rate dependent (see e.g. Labud et al.[1], and Figure 4). Therefore an effect with a steep increase in rate like the resonant electron tunnelling must be responsible for the observed excitonic peaks. The spatially indirect tunnel recombination rate $\Gamma_{\widetilde{m}\to m}^{\mathrm{e}^-,\,\mathrm{indirect}}$ must thus be orders of magnitude weaker than the resonant electron tunnel rate. The reason for this is most probably the weaker wavefunction overlap of an electron in the back contact with a heavy hole state in the QD and the influence of the dipole matrix element compared to the overlap with an electron state in the QD [9].

## Section II d): Solution of the master equation: no illumination

For the non-illuminated case, the master equation is exemplarily calculated for charges up to two electrons (0, -1, -2). The three electron charge level (-3) is split $\approx 45$ meV away and thus not considered. Due to the quasi equilibrium we start with neighbouring charge configurations only.

$$m = 0 \Rightarrow \quad \Gamma^{e^-}_{-1 \to 0} p_{-1} - \Gamma^{e^-}_{0 \to -1} p_0 = 0$$

$$m = -1 \Rightarrow \quad \Gamma^{e^-}_{0 \to -1} p_0 + \Gamma^{e^-}_{-2 \to -1} p_{-2} - (\Gamma^{e^-}_{-1 \to 0} + \Gamma^{e^-}_{-1 \to -2}) p_{-1} = 0 \quad (12)$$

$$m = -2 \Rightarrow \quad \Gamma^{e^-}_{-1 \to -2} p_{-1} - \Gamma^{e^-}_{-2 \to -1} p_{-2} = 0$$

$$p_0 + p_{-1} + p_{-2} = 1$$

These four equations can now be solved for the charge state probabilities

$$p_0 = \frac{1}{1 + \frac{\Gamma^{e^-}_{0 \to -1}}{\Gamma^{e^-}_{-1 \to 0}} + \frac{\Gamma^{e^-}_{0 \to -1}}{\Gamma^{e^-}_{-1 \to 0}} \frac{\Gamma^{e^-}_{-1 \to -2}}{\Gamma^{e^-}_{-2 \to -1}}}$$

$$p_{-1} = \frac{1}{1 + \frac{\Gamma^{e^-}_{-1 \to 0}}{\Gamma^{e^-}_{0 \to -1}} + \frac{\Gamma^{e^-}_{-1 \to -2}}{\Gamma^{e^-}_{-2 \to -1}}} \quad (13)$$

$$p_{-2} = \frac{1}{1 + \frac{\Gamma^{e^-}_{-2 \to -1}}{\Gamma^{e^-}_{-1 \to -2}} + \frac{\Gamma^{e^-}_{-1 \to 0}}{\Gamma^{e^-}_{0 \to -1}} \frac{\Gamma^{e^-}_{-2 \to -1}}{\Gamma^{e^-}_{-1 \to -2}}}$$

At low temperatures we can safely restrict ourselves to transitions between one charge state towards another. Thus, these charge transition points (and also the peaks in the measurements) occur at $p_0 = p_{-1} = 1/2$ and $p_{-1} = p_{-2} = 1/2$. For the first transition $(0 \to -1)$ we can assume the term $\Gamma^{e^-}_{-1 \to -2}/\Gamma^{e^-}_{-2 \to -1}$ to be zero as the nominator will be very small (no tunnelling in for a second electron) and the denominator becomes large (tunnelling out rates for a two electron charge state will be large). Thus, the first charge transition will occur at

$$\Gamma^{e^-}_{-1 \to 0} = \Gamma^{e^-}_{0 \to -1}. \quad (14)$$

Within equivalent arguments the second charge transition $(-1 \rightarrow -2)$ becomes

$$\Gamma^{e^-}_{-2 \rightarrow -1} = \Gamma^{e^-}_{-1 \rightarrow -2}. \tag{15}$$

These equilibria are used in the main text for the non-illuminated measurement s-peak shifts. A derivation of the exciton peak is analogue and yields

$$\Gamma^{e^-}_{1 \rightarrow 0} = \Gamma^{h, \text{gen}}_{0 \rightarrow 1}. \tag{16}$$

## Section II e): Hole generation and electron tunnelling rate approximation

A precise expression for the hole generation rate is not trivial, as it depends on the photon number flux entering the Schottky diode, the absorption coefficients in the QDs and wetting layer and electron-hole pair dissociation rate.

We now want to motivate our suggestion that the hole generation rate is at a very low level on the scale in the plot of Figure 3. For our sample, the tunnel rate into and out of the non-hole-charged electron $s_1$ state is on the order of $1$ MHz (In Luyken et al.[4], a method for estimating the tunnel rate is given.). As the exciton related peaks already get significantly suppressed at $1$ kHz at an LED current of $I_{\text{LED}} = 2$ mA (also see supplemental material of the paper by Labud et al. [1], where the $X^0$ peak is already suppressed by 50 % at $1$ kHz), we can conclude, that the experimentally observed tunnel in rate (equal hole generation rate) is small compared to the step height of the first step in Figure 3 (maximum theoretical tunnel in rate). This is mainly due to the limiting factor of the hole generation rate (equation (16) and next section). We conclude from the best fit of our model to the experimental data (Figure 3 (b), main text), that our hole generation rate is at $\approx 400$ Hz.

## Section III: Hole generation rate variation

This directly leads to an experiment, where we varied systematically the LED current, increasing the hole generation rate (Figure 4).

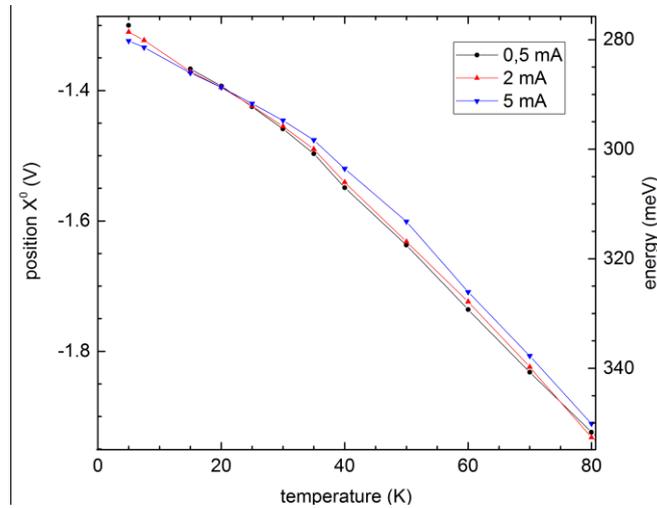

Figure 4: $X^0$ peak position as a function of temperature for various LED-illumination currents with connecting lines as guide to the eye. At low temperatures up to 15 K we see a larger shift due to heating of the sample by the LED power dissipation. At intermediate temperatures, a less pronounced peak shift is visible for larger LED-currents, in agreement with our model. At even higher temperatures the situation becomes more complicated. But it still holds, that increasing the hole generation rate by stronger illumination does not dramatically change the peak position.

At first, we observe an unexpected shift towards more reverse biases at low temperatures. This is not observed in the $T = 4.2$ K study by Labud et al. [1], where the measurements are performed in a helium bath cryostat. We attribute the shift at low temperatures in our experiment to heating of the sample at higher LED currents at low temperatures, also monitored by the sample heater output power. This is the case for our experiment as the thermal conductivity of the thermal contact provided by the stainless steel cables in our experiment is not very high in contrast to the experiment in the liquid helium bath cryostat by Labud et al. where thermal conductivity is provided by helium as an exchange gas and copper wires are used. At higher temperatures, a clear shift to higher forward gate voltages is observed. Heating from power dissipation of the LED can now be neglected, as a) the LED voltage is lower and thus the quantum efficiency higher, b) the heat capacity and c) thermal conductivity is higher for higher temperatures in our setup. The observed shift with increasing LED current is well explained in the frame of our model, as the function (Equation 15) is now evaluated at a higher tunnel rate (see also Figure 3) due to the higher hole generation rate.

## Section IV: Broadening of the Peaks

Increasing the temperature, the peaks in our C(V)-measurements get broadened. Can this be explained by apparently random tunnelling electrons below and above the resonance condition? Usually, the Lock-In amplifier should be able to filter such random processes out. However, the probability of this processes changes with the gate voltage and thus also with the AC-modulation voltage, adding a signal at the demodulator frequency, even not in resonance with the QD levels.